\documentclass[]{spie}  

\usepackage{amsmath,amsfonts,amssymb}
\usepackage{graphicx}
\usepackage[colorlinks=true, allcolors=blue]{hyperref}
\usepackage{float}

\title{Generating electron beam lithography write parameters from the FORTIS holographic grating solution}

\author[a]{Mackenzie Carlson}
\author[a]{Stephan McCandliss}
\author[b]{Randall McEntaffer}
\author[b]{Fabien Gris\'e}
\author[c]{Nicholas Kruczek}
\author[c]{Brian Fleming}
\affil[a]{Johns Hopkins University, Baltimore, USA}
\affil[b]{Pennsylvania State University, University Park, USA}
\affil[c]{University of Colorado, Boulder, USA}

\begin{document} 
\maketitle

\begin{abstract}
The Far-UV Off Rowland-circle Telescope for Imaging and Spectroscopy (FORTIS) has been successful in maturing technologies for carrying out multi-object spectroscopy in the far-UV, including: the successful implementation of the Next Generation of Microshutter Arrays; large-area microchannel plate detectors; and an aspheric “dual-order” holographically ruled diffraction grating with curved, variably-spaced grooves with a laminar (rectangular) profile. These optical elements were used to construct an efficient and minimalist “two-bounce” spectro-telescope in a Gregorian configuration. However, the susceptibility to Lyman alpha (Ly$\alpha$) scatter inherent to the dual order design has been found to be intractably problematic, motivating our move to an “Off-Axis” design. OAxFORTIS will mitigate its susceptibility to Ly$\alpha$ by enclosing the optical path, so the detector only receives light from the grating. The new design reduces the collecting area by a factor of 2, but the overall effective area can be regained and improved through the use of new high efficiency reflective coatings, and with the use of a blazed diffraction grating. This latter key technology has been enabled by recent advancements in creating very high efficiency blazed gratings with impressive smoothness using electron beam lithography and chemical etching to create grooves in crystalline silicon. Here we discuss the derivation for the OAxFORTIS grating solution as well as methods used to transform the FORTIS holographic grating recording parameters (following the formalism of Noda et al.1974a,b), into curved and variably-spaced rulings required to drive the electron beam lithography write-head in three dimensions. We will also discuss the process for selecting silicon wafers with the proper orientation of the crystalline planes and give an update on our fabrication preparations. 
\end{abstract}


\section{INTRODUCTION}
\label{sec:intro}  

FORTIS is a “two bounce” Gregorian telescope with a diffractive secondary. It was designed to carry out multi-object spectroscopy in the far-UV and has completed four launches. A reoccurring issue from each launch has been the geocoronal Lyman Alpha (Ly$\alpha$) scatter off of the optical train assembly (OTA) that result in high count rates. Attempts were made during each consecutive flight to minimize this Ly$\alpha$ scatter with improved baffles. Despite the success in reducing the Ly$\alpha$ scatter by two orders of magnitude in comparison to its first flight, the background rate is still approximately 100 times higher than the detector dark rate.

\begin{figure} [H]
   \begin{center}
   \includegraphics[height=7.5cm,trim={3.5cm 2.5cm 2.5cm 0cm},clip]{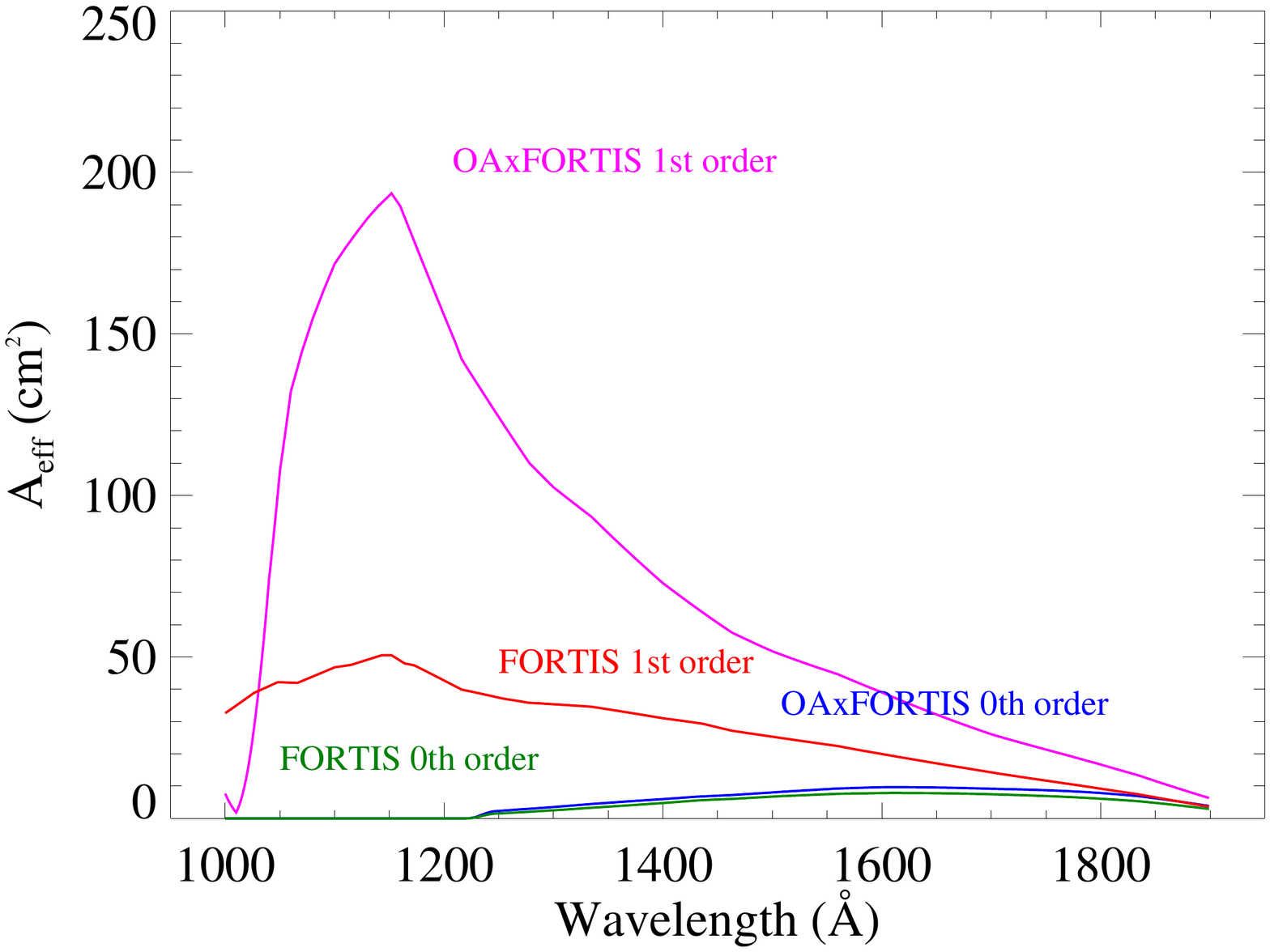}
   \includegraphics[height=6.2cm]{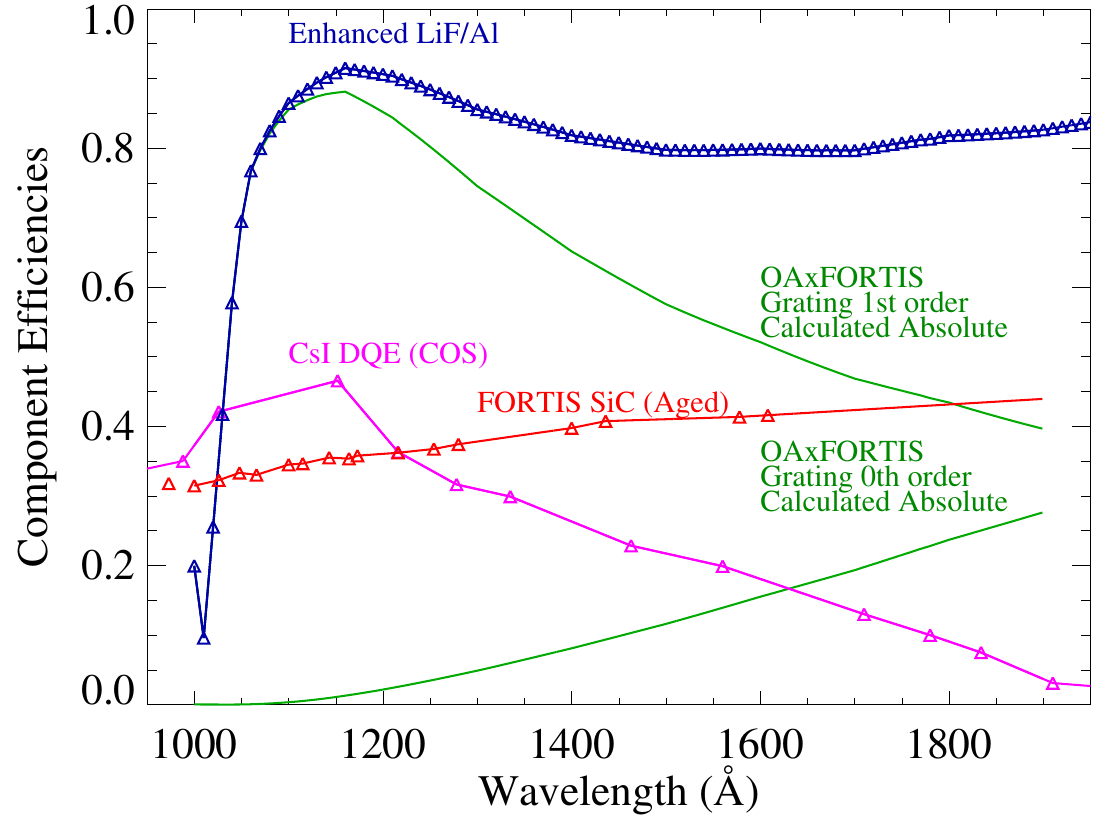}
   \end{center}
   \caption[example] 
   { \label{effarea} 
(Right) - Effective area as a function of wavelength for OAxFORTIS and FORTIS 0th and 1st orders. (Left)~-~Component efficiencies as a function of wavelength for OAxFORTIS for LiF/Al, CsI detective quantum efficiency, and absolute in 0th and 1st orders, as well as FORTIS for SiC. }
   \end{figure}

Similar to FORTIS, the new OAxFORTIS is a f/10 Gregorian spectro-telescope with a parabolic primary and diffractive elliptic secondary. To remedy the scattering issue, the OTA is being redesigned to include enclosure of the optical path so the detector only receives light from the grating. This requires blocking half the aperture; however, recent innovations in mirror coating and grating fabrication technologies will allow OAxFORTIS to have many times the effective area of FORTIS, as seen in Figure (\ref{effarea}). Among these innovations and technologies are blazed diffraction gratings fabricated via electron beam lithography (EBL),\cite{McEntaffer:2013,Grise:2020,Beasley:2019,McCoy:2020} enhanced LiF/Al mirror coatings,\cite{Fleming:2017,Quijada:2017} and a 3rd generation microshutter array (MSA-3G).\cite{Kim:2021,Kutyrev:2020} The translation of holographic to mechanical grating rulings and the design of the blazed diffraction grating will be the focus of this paper.
   
\begin{figure} [H]
   \begin{center}
   \begin{tabular}{c} 
   \includegraphics[height=11cm,trim={0 1cm 0 0cm},clip]{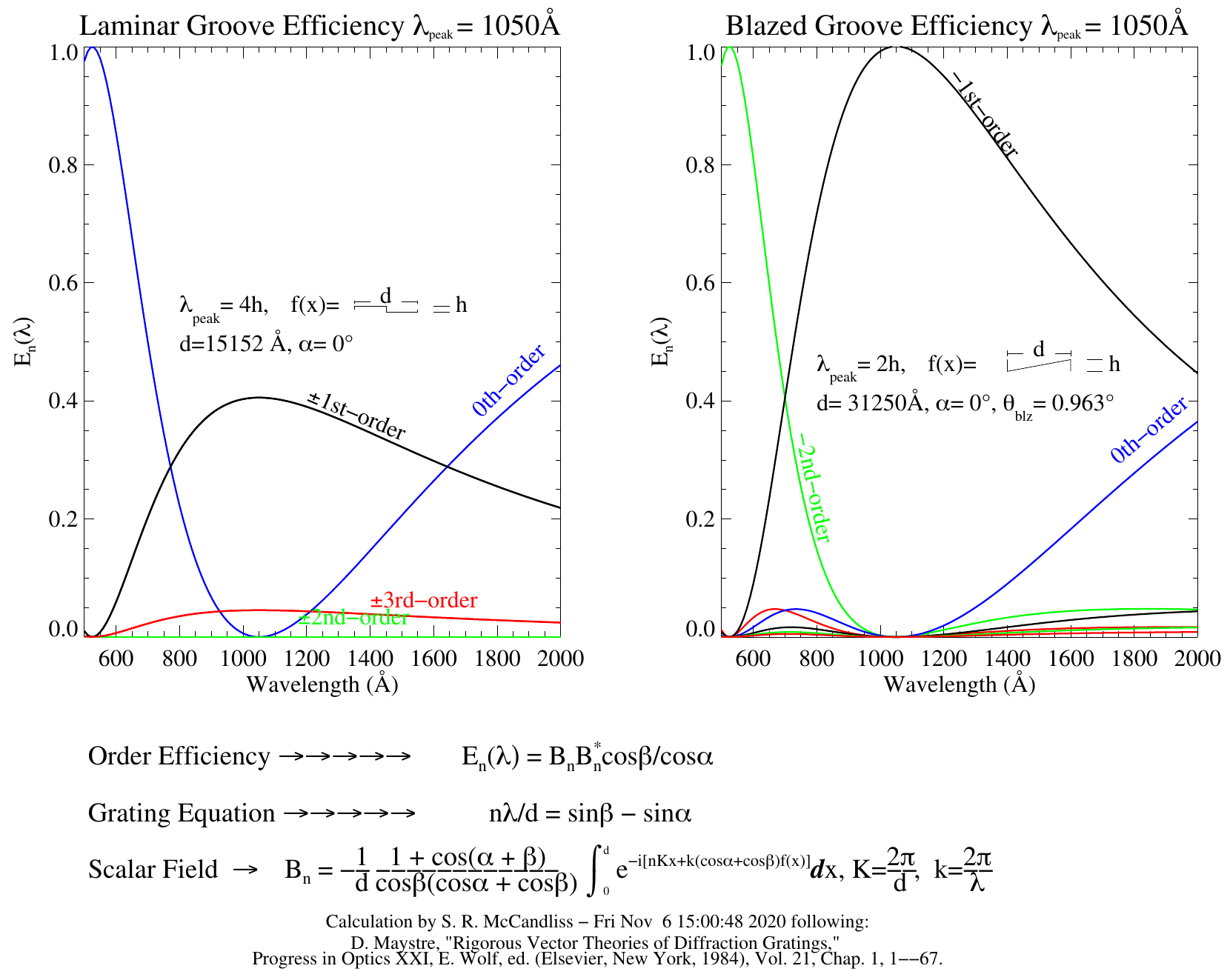}
   \end{tabular}
   \end{center}
   \caption[example] 
   { \label{grateff} 
A comparison of laminar and idealized blazed groove efficiencies for a peak efficiency at 1050Å.}
   \end{figure}

 The original version of FORTIS used a holographically ruled diffractive secondary with a triaxial-elliptic figure to reduce astigmatism in the spectral channels. This secondary has a laminar groove profile, and provides both an imaging channel and high throughput spectroscopy in dual orders for multiple objects within a (0.5°)$^2$ field-of-view (FOV).\cite{McCandliss10} However, a blazed grating can achieve a theoretical groove efficiency of 100\%, which is higher than the ($\sim$80\%) sum of the ± 1st orders of a laminar grating and places all the light on a single detector, eliminating the background penalty from co-adding two channels. Figure (\ref{grateff}) shows a comparison of these groove efficiencies for a grating with a peak efficiency at 1050Å. A potential difficulty is the requirement for a very low blaze angle ($\sim$1°). In case of issues arising with the blazed grating, we are also having a laminar grating produced with the same fabrication process and figure. Ruling parameters for a new design were derived recently using Noda et al.\cite{Noda74a,Noda74b} aberration correction formalism. This new solution was found to visibly minimize the effects of astigmatism in the 1st order across a wavelength range of 900-1800Å as seen in Figure~(\ref{aberr}), which is partially due to the reduced aperture, but also because the single order solution controls astigmatism more effectively. 
 
 Holographic rulings, such as those on the FORTIS secondary, are the result of monochromatic beams from two coherent light sources creating a pattern of interference fringes on the photoresist-coated surface of a grating blank.\cite{Noda74b} While this technique is convenient and inexpensive, a new method of mechanically ruling gratings via electron beam lithography can result in impressive groove smoothness.\cite{McCoy:2020,Grise:2020} The extremely precise (laser interferometer controlled) e-beam tool used to write the rulings into an electron beam resist on the grating's surface requires a map of the desired three-dimensional groove pattern to trace. Here we will present the derivation for the dual order and blazed grating solutions (Section 2) and the method to translate these into polynomial coordinates suitable for driving the e-beam tool (Section 3) followed by the fabrication steps and future work needed to finish the OAxFORTIS grating (Section 4). The goal of this process is to develop the steps needed to consistently produce holographically derived variably-spaced curved groove gratings with tunable blaze angles on aspheric (free-form) substrates.

\section{GRATING DESIGN DERIVATION}
\label{sec:grating} 
The derivation of ruling parameters for a diffraction grating begins with the light-path function $F$ expansion in terms of the aberration coefficients from Noda et al. such that,
\begin{equation}
\label{Fijk}
F_{ijk}=M_{ijk} + m\frac{\lambda}{\lambda_0}H_{ijk},
\end{equation}
where the subscripts $\textit{ijk}$ are the exponents of the $\textit{xyz}$ positions on the grating surface,\cite{Wilkinson} m is the spectral order, $\lambda$ is the use wavelength, $\lambda_0$ is the wavelength used to record the holographic pattern, and the $M_{ijk}$ and $H_{ijk}$ terms are the mechanical and holographic components, respectively.\cite{Noda74b}  The $F_{200}$ and $F_{020}$ terms encode for meridional and sagittal astigmatism.\cite{Tsonev} Their components are:
\begin{equation}
\label{eq:M200}
M_{200} = \frac{cos^2\alpha}{r} + \frac{cos^2\beta}{r'} - 2a_{20}(cos\alpha+cos\beta),
\end{equation}
\begin{equation}
\label{H200}
H_{200} = \frac{cos^2\gamma}{r_C} - \frac{cos^2\delta}{r_D} - 2a_{20}(cos\gamma-cos\delta),
\end{equation}
\begin{equation}
\label{M020}
M_{020} = \frac{1}{r} + \frac{1}{r'} - 2a_{02}(cos\alpha+cos\beta)   ,
\end{equation}
\begin{equation}
\label{H020}
H_{020} = \frac{1}{r_C} - \frac{1}{r_D} - 2a_{02}(cos\gamma-cos\delta) ,
\end{equation}
where $r$ is the object distance ($s$), $r'$ is the 0th order image distance ($s_m$), $r_C$ and $r_D$ are the recording legs, $\delta$ and $\gamma$ are the recording angles defined with respect to the grating normal, $\alpha$ and $\beta$ are the angles of incidence and diffraction for the use geometry, and the $a_{ij}$ terms are the elliptical constants listed in Table (\ref{table}).\cite{Noda74a} We also need the equation for the effective grating constant, $\sigma$,
\begin{equation}
\label{sigma}
\frac{\lambda_0}{\sigma} = sin\delta - sin\gamma,
\end{equation}
as well as the grating equation,
\begin{equation}
\label{grateq}
\frac{m\lambda_0}{\sigma} = sin\alpha + sin\beta,
\end{equation} 
as these define and constrain the diffraction grating rulings.

FORTIS has a requirement to minimizing astigmatism over as wide an angular field as possible. Wide field spectroscopy in off-Rowland circle configurations with straight, uniformly spaced rulings can not yield the required spatial resolution; therefore, the design for FORTIS involved curved, variably-spaced rulings. The dual order, off-Rowland circle grating design for FORTIS utilizes a fully mechanical solution to minimize astigmatism equally in both positive and negative orders. The holographic components of Equation (\ref{Fijk}) become zero by enforcing a symmetrical recording geometry such that $\gamma$~=~-$\delta$ and $r_C$~=~$r_D$ so that $H_{200}$~=~0~=~$H_{020}$. We are left with just the mechanical terms to manipulate and horizontal and vertical focal curves are then obtained.\cite{Noda74a} These are written as,
\begin{equation}
\label{F200FORT}
0 = M_{200} = \frac{1}{r} + \frac{cos^2\beta}{r'} - 2a_{20}(1+cos\beta)   ,
\end{equation}
\begin{equation}
\label{F020FORT}
0 = M_{020} = \frac{1}{r} + \frac{1}{r'} - 2a_{02}(1+cos\beta) ,
\end{equation}
since we set $\alpha$=0. The ratio of Eqs. (\ref{F200FORT}) and (\ref{F020FORT}) is then found which can be arranged to reach a solution such that,
\begin{equation}
\label{FORTsoln}
\frac{m + cos^2\beta_c}{m +1} = \frac{b^2}{a^2},
\end{equation}
where m is now the magnification ratio from prime focus to secondary focus, and $a$ and $b$ are the grating's (triaxial) elliptical axis constants, listed in Table (\ref{table}).

\begin{figure} [H]
   \begin{center}
   \begin{tabular}{c} 
   \includegraphics[height=10cm,trim={0 0.6cm 0 0},clip]{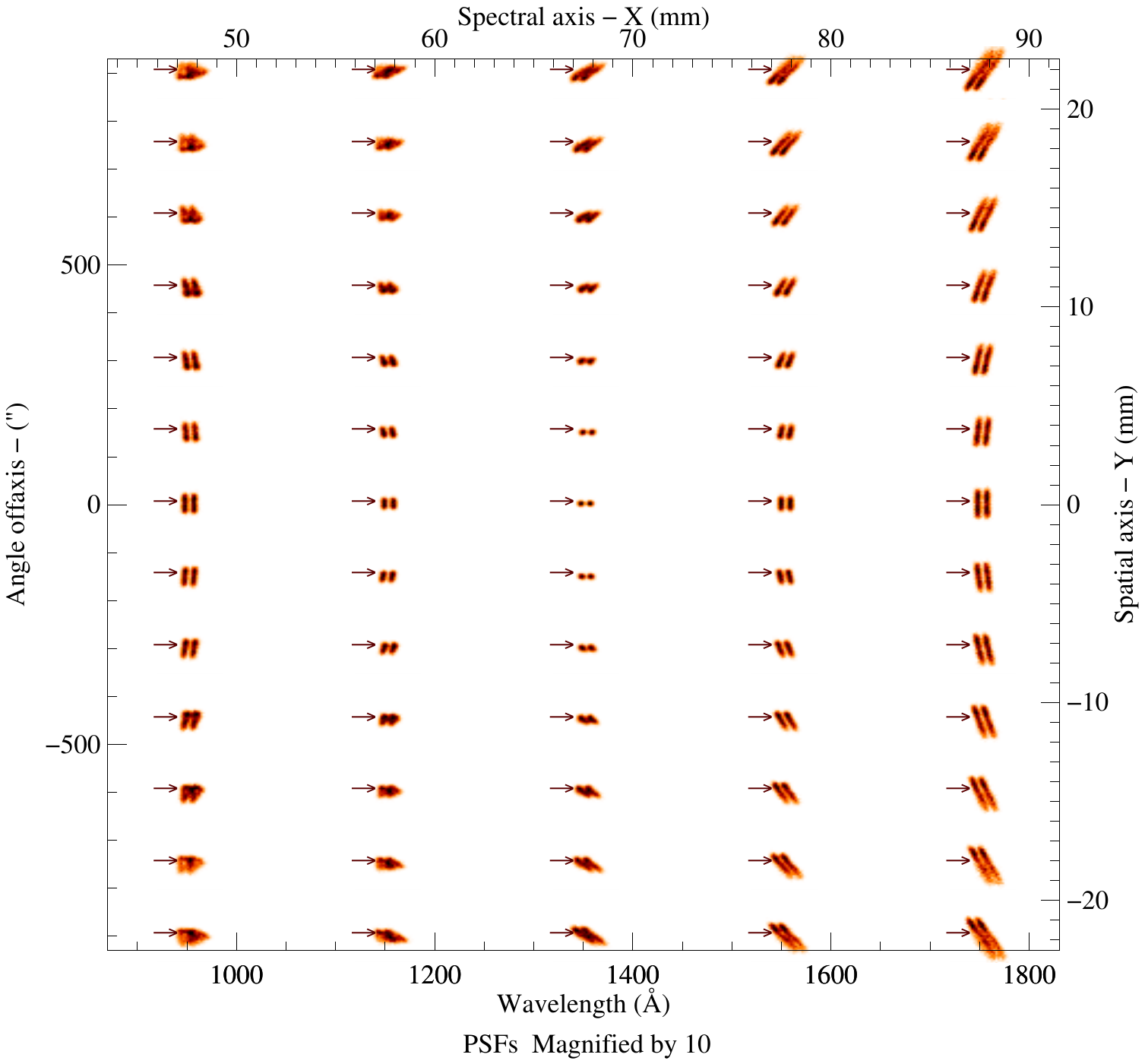}\\
   \includegraphics[height=10.5cm,trim={0 0cm 0 0cm},clip]{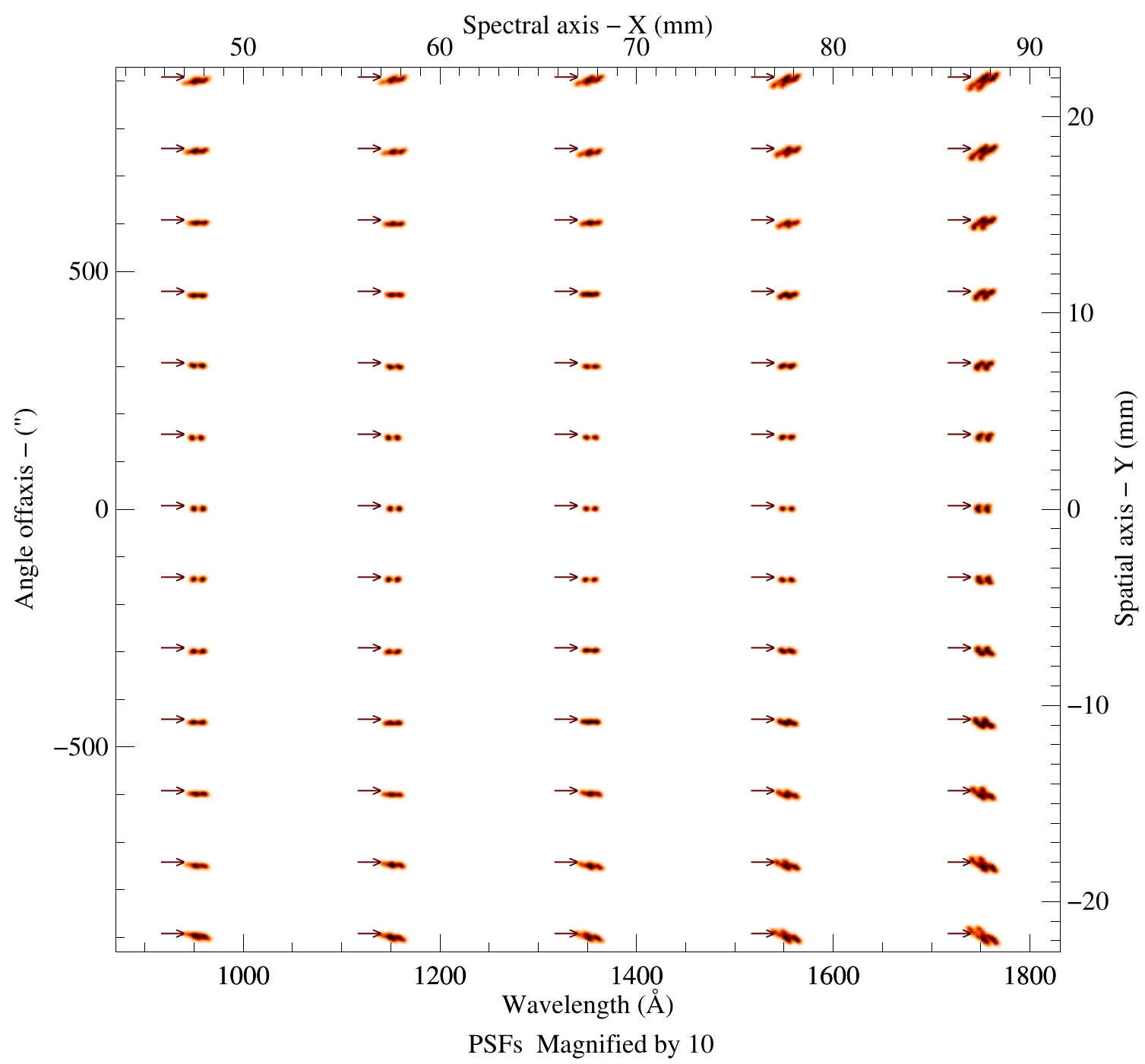}\\ 
   \end{tabular}
   \end{center}
   \caption[example] 
   { \label{aberr} 
(Top) - Point spread function (PSF) as a function of wavelength and off-axis angle in 1st order for FORTIS dual order mechanical design. (Bottom) - PSF for OAxFORTIS blazed solution with zero order.}
   \end{figure}

The mechanical solution reduces astigmatism equally in the first orders, at the expense of introducing astigmatism in the zeroth order; however, a simple cylindrical doublet is used to correct this aberration. OAxFORTIS only requires astigmatism minimization in a singular order, but still requires an acceptable zero order image to aid in focal location. The blazed solution minimizes astigmatism by balancing the mechanical and holographic components. The $\gamma$-$\delta$ symmetry is dropped and the $F_{200}$=0 and $F_{020}$=0 curves are set to equal each other. Again with $\alpha$=0 and recording legs $r_C$=$r_D$, these are written as,
\begin{equation}
\label{F200OAx}
0 = F_{200} = \frac{1}{r} + \frac{cos^2\beta}{r'} - 2a_{20}(1+cos\beta) -  m\frac{\lambda}{\lambda_0}\left [\frac{cos^2\gamma-cos^2\delta}{r_C} - 2a_{20}(cos\gamma-cos\delta)\right ]   ,
\end{equation}
\begin{equation}
\label{F020OAx}
0 = F_{020} = \frac{1}{r} + \frac{1}{r'} - 2a_{02}(1+cos\beta) - 2m\frac{\lambda}{\lambda_0}a_{20}(cos\gamma-cos\delta) .
\end{equation}
By applying Equation (\ref{sigma}) as a constraint on the relationship for $\gamma$ and $\delta$,
a solution for $\gamma$ can be derived,
\begin{equation}
\label{gamma}
\gamma = sin^{-1}\left ( -\frac{\lambda_0}{2\sigma} - \frac{\sigma  sin^2\beta_c}{2\lambda_c} \right ),
\end{equation}
with $\lambda_c$ as the central wavelength for astigmatism correction and $\beta_c$ being the angle corresponding to it. The $\delta$ recording angle can be found using Eqs.~(\ref{sigma}) and (\ref{gamma}). Comparing point spread function plots in Figure (\ref{aberr}) for both grating designs, it is clear that this new blazed grating solution is a superior choice for minimizing astigmatism in OAxFORTIS. Despite this, there remains visible aberrations in the edges of the FOV. In the future, we will explore optimization of this grating solution for our wide FOV to further minimize the lingering astigmatism. In the next section, we will derive groove polynomials for both grating designs described in this section.

\begin{table}[H]
\caption{Geometry and Recording Parameters for Old and New Gratings} 
\label{table}
\begin{center}       
\begin{tabular}{|l|l|} 
\hline
\rule[-1ex]{0pt}{3.5ex} \bf Parameter & \bf Value  \\
\hline
\rule[-1ex]{0pt}{3.5ex}  Triaxial Ellipsoidal x-axis & a = 698.77 mm   \\
\hline
\rule[-1ex]{0pt}{3.5ex}  Triaxial Ellipsoidal y-axis & b = 698.67 mm  \\
\hline
\rule[-1ex]{0pt}{3.5ex}  Triaxial Ellipsoidal z-axis & c = 937.50 mm  \\
\hline
\rule[-1ex]{0pt}{3.5ex}  Ellipsoidal x and y axis & a = b = 698.77 mm  \\
\hline 
\rule[-1ex]{0pt}{3.5ex}  Ellipsoidal z axis & c = 937.50 mm  \\
\hline 
\rule[-1ex]{0pt}{3.5ex}  x-axis Parameter & $a_{20} = c/2a^2$  \\
\hline 
\rule[-1ex]{0pt}{3.5ex}  y-axis Parameter & $a_{02} = c/2b^2$  \\
\hline 
\rule[-1ex]{0pt}{3.5ex}  Image Distance & $s_{m}$ = 1562.50 mm  \\
\hline 
\rule[-1ex]{0pt}{3.5ex}  Object Distance & $s$ = 312.50 mm  \\
\hline 
\rule[-1ex]{0pt}{3.5ex}  Stigmatic Wavelength & $\lambda_c$ = 1300 Å  \\
\hline 
\rule[-1ex]{0pt}{3.5ex}  Output Angle for $\lambda_c$ & $\beta_c$ = 2.38°  \\
\hline 
\end{tabular}
\end{center}
\end{table}

\section{GROOVE PATTERN SOLUTION}
\label{sec:solution}
 
\subsection{Grating Design Theory}
Here we present a method to translate the holographic grating prescription for variable spaced rulings into unique polynomial equations. Our origin is the vertex of the concave grating with the z-axis normal to the grating at 0.\footnote{z-x is the dispersion plane and z-y is perpendicular to the plane of dispersion. This differs from Noda et al.} There are two coherent point sources, C and D, which are used for recording interference fringes on the grating.\cite{Noda74b} Additionally, points A(x,y,z), P(w,l,$\xi$), and B(x',y',z') can be described respectively as a self-luminous point in the entrance slit, a point on the nth groove counted from the groove passing through the origin, and the point of a ray with wavelength $\lambda$  diffracted from P in the mth order.\cite{Noda74b} For the ray $APB$, the light-path function F is defined by,
\begin{equation}
\label{F}
F=\left\langle AP \right\rangle + \left\langle PB \right\rangle + nm\lambda   ,
\end{equation}
where $\left\langle AP \right\rangle$ and $\left\langle PB \right\rangle$ are the distances between the respective points and n is given by,
\begin{equation}
\label{nlam}
n\lambda_{o}=[\left\langle CP \right\rangle - \left\langle DP \right\rangle] - [\left\langle CO \right\rangle - \left\langle DO \right\rangle]  ,
\end{equation}
with $\lambda_{o}$ as the recording wavelength.\cite{Noda74b}

We apply their technique first to our triaxial ellipse dual order grating. To begin, we set $[\left\langle CO \right\rangle~-~\left\langle DO \right\rangle]$=0 in Equation~(\ref{nlam}) for simplicity without compromising generality and,
\begin{equation}
\label{CP}
\left\langle CP \right\rangle = \sqrt{(s_{m}cos\gamma -\xi )^{2}+(s_{m}sin\gamma -w)^{2}+l^{2}},
\end{equation}
\begin{equation}
\label{DP}
\left\langle DP \right\rangle = \sqrt{(s_{m}cos\delta -\xi )^{2}+(s_{m}sin\delta -w)^{2}+l^{2}},
\end{equation}
where the recording legs $r_C=r_D$ are equal to $s_{m}$, the coordinates $\textit{w}$ and $\textit{l}$ are the x and y directions for the grating geometry, and $\xi$ is the depth coordinate.\cite{Noda74a} The equation used to represent the triaxial ellipsoid shape of the dual order grating, when rearranged to be equal to $\xi$, is
\begin{equation}
\label{z}
\xi = c - \sqrt{c^{2}+\left (\frac{wc}{a}  \right )^{2}+\left (\frac{lc}{b} \right)^{2}},
\end{equation}
where $\textit{a}$, $\textit{b}$, and $\textit{c}$ are positive axis constants. The equation we use in our blazed grating solution is the same, but now with $\textit{a}$ = $\textit{b}$ so the ellipsoidal figure is no longer triaxial. The axis constants for both gratings can be found in Table (\ref{table}). In addition to this change in $\xi$ as a result of the new grating solution, Eqs.~(\ref{CP}) and (\ref{DP}) are also slightly different due to the recording legs being equal to $s_{m}/cos\beta_{c}$.

At this point the set of equations needed is defined and we are ready to solve this groove problem. The final solution is obtained by solving this for the $\textit{w}$, $\textit{l}$ coordinates for each groove number, $\textit{n}$, which make up the pattern of grooves to be etched onto our grating. Next, we will outline two different methods we have applied to help validate the solution.

   \begin{figure} [H]
   \begin{center}
   \begin{tabular}{c} 
   \includegraphics[height=12cm,trim={2cm 2cm 2cm 2cm},clip]{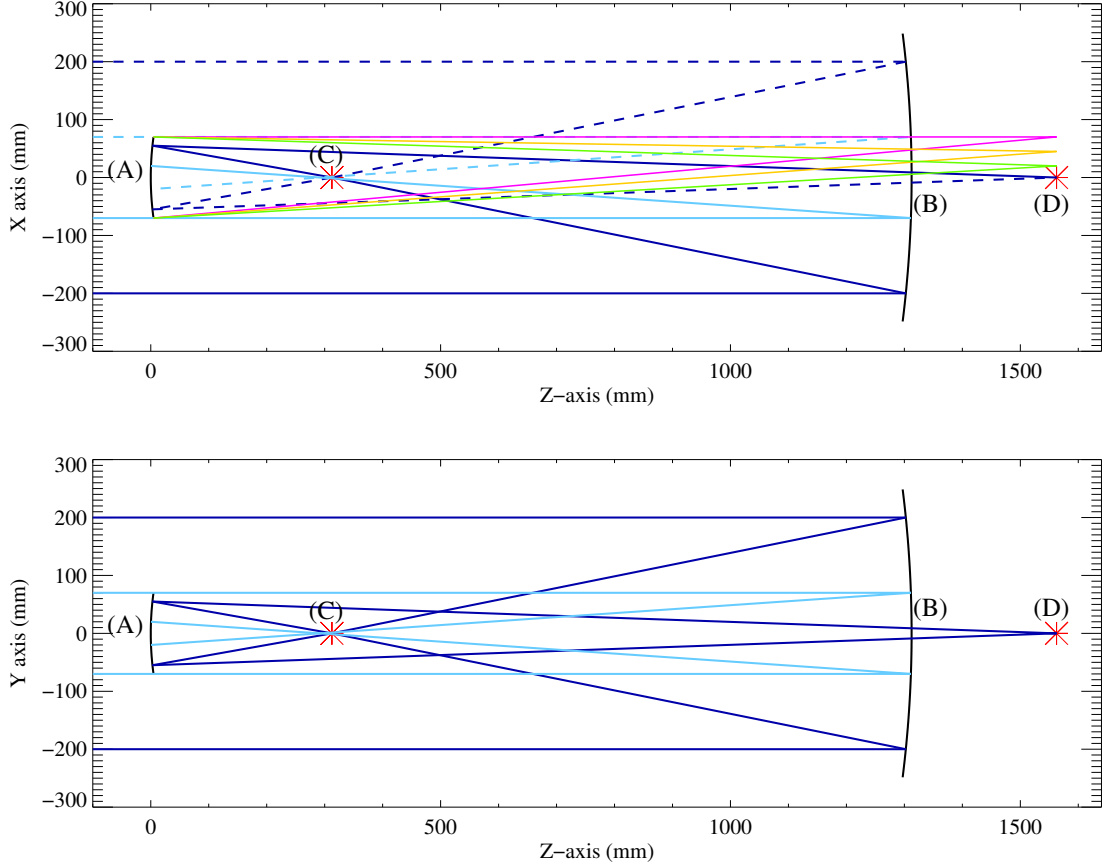}
   \end{tabular}
   \end{center}
   \caption[example] 
   { \label{diagram} 
Optical path diagram for OAxFORTIS in the dispersion plane (Top) and orthogonal plane (Bottom). Dashed lines represent light cut off by the halved aperture and solid lines are actual light rays. The light blue lines are the central obscurations from the pupil. The pink, yellow, and green rays are the dispersion rays. (A) – Vertex of the diffractive secondary mirror. (B) - Vertex of the primary mirror. (C) – Prime Focus; The distance AC = $s$. (D) - Secondary Focus; The distance AD = $s_m$ = $s*m$ where m is the magnification.  }
   \end{figure} 

\subsection{Shotgun Method}
One method for mapping the grooves is a ``shotgun" solution. The goal was to find the $\textit{w}$ and $\textit{l}$ coordinates where n is approximately an integer. To start this, matrices of randomly selected $w$ and $l$ coordinates are created. Using Eqs.~(\ref{nlam}) and (\ref{z}) and by looping through every 500th groove, indices in the $w$ and $l$ coordinates are extracted where the absolute value of the difference between -$\textit{n}$ and this looping variable is less than 0.05. Then, the $\textit{w}$ and $\textit{l}$ values corresponding to these indices are selected, providing a theoretically complete set of coordinates for each groove. As seen in Figure (\ref{old}), these grooves now seem to take on a parabolic shape. To test the accuracy of this observation, we conducted polynomial fits through which we found that each groove – excluding $\textit{n}$ = 0 which is a straight line – can nearly perfectly be described as a parabola. 

\subsection{Exact Method}
An exact solution is available via an algebraic method to extract values for the groove polynomials. By inserting Equation~(\ref{z}) and the known constants into Eqs.~(\ref{nlam}), (\ref{CP}), and (\ref{DP}), we work towards a simplified polynomial equation equal to 0 with multiple terms grouped by powers of $w$ up to $\textit{w}^4$. The final result was of the form $0 = Aw^4 + Bw^3 + Cw^2 + Dw + Const$, with $\textit{A}$, $\textit{C}$, and $\textit{Const}$ consisting of sums of multiple terms dependent on $\textit{n}$ and $\textit{l}$, with $\textit{B}$ and $\textit{D}$ canceling out. This polynomial is looped through the full range of $\textit{l}$ coordinate values, as well as each 500th groove, leaving just the $\textit{w}$ variable remaining for each loop. Then, the $\textit{w}$ root value of this polynomial is found at each $\textit{l}$ coordinate for each of the grooves, providing a collection of coordinates.
   \begin{figure} [H]
   \begin{center}
   \begin{tabular}{c} 
   \includegraphics[height=9cm]{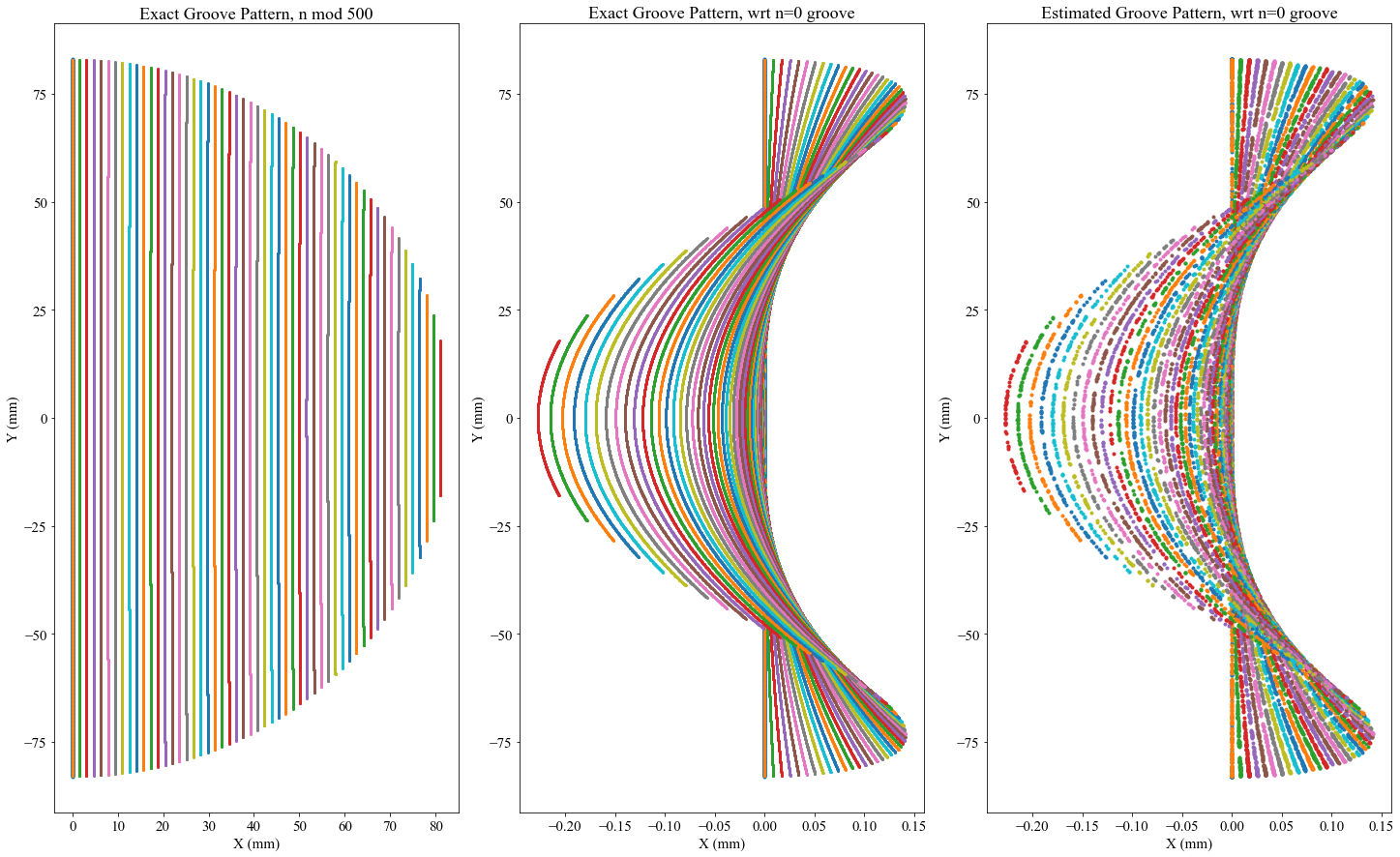}
   \end{tabular}
   \end{center}
   \caption[example] 
   { \label{old} 
(Left) – Variable spaced curved groove solution for half of grating, shown mod 500. (Middle) - Algebraically derived groove pattern for original grating solution with respect to the effective width. (Right) – Estimated groove pattern with respect to the effective width. }
   \end{figure} 

The above gives the results for the $\textit{n}$ mod 500 pattern, as shown in Figure~(\ref{old}). To force this groove pattern to be respective to the effective groove width at $\textit{n}$ = 0, a correction factor is applied to $\textit{w}$ so that the groove pattern seen in Figure (\ref{old}) is graphed as $\textit{l}$ vs. $\textit{w}$ – $\textit{n/d}$ where $\textit{d}$ is the ruling density. This results in a variably spaced groove pattern and each groove has the same parabolic shape described in Section 3.2. As seen in Figure~(\ref{old}), the shotgun solution and the algebraically derived solution appear to match very well, bringing confidence to our results. Given the higher accuracy of the algebraic method, and the fact that there are no potential gaps in the pattern, as there can be with the shotgun solution, we will be using this new method for our gratings. However, it is very useful to have the shotgun method available to confirm the groove pattern found via this exact method.

\subsection{Blazed Solution}
The exact method described above was also used for the blazed grating with zero order design. Applying the new geometry and ruling parameters described in Section 2 to this method, a similar final polynomial equation is extracted, this time with powers of $w$ up to $w^8$. The resulting groove pattern is imaged below in Figure (\ref{new}).
   \begin{figure} [H]
   \begin{center}
   \begin{tabular}{c} 
   \includegraphics[height=9cm]{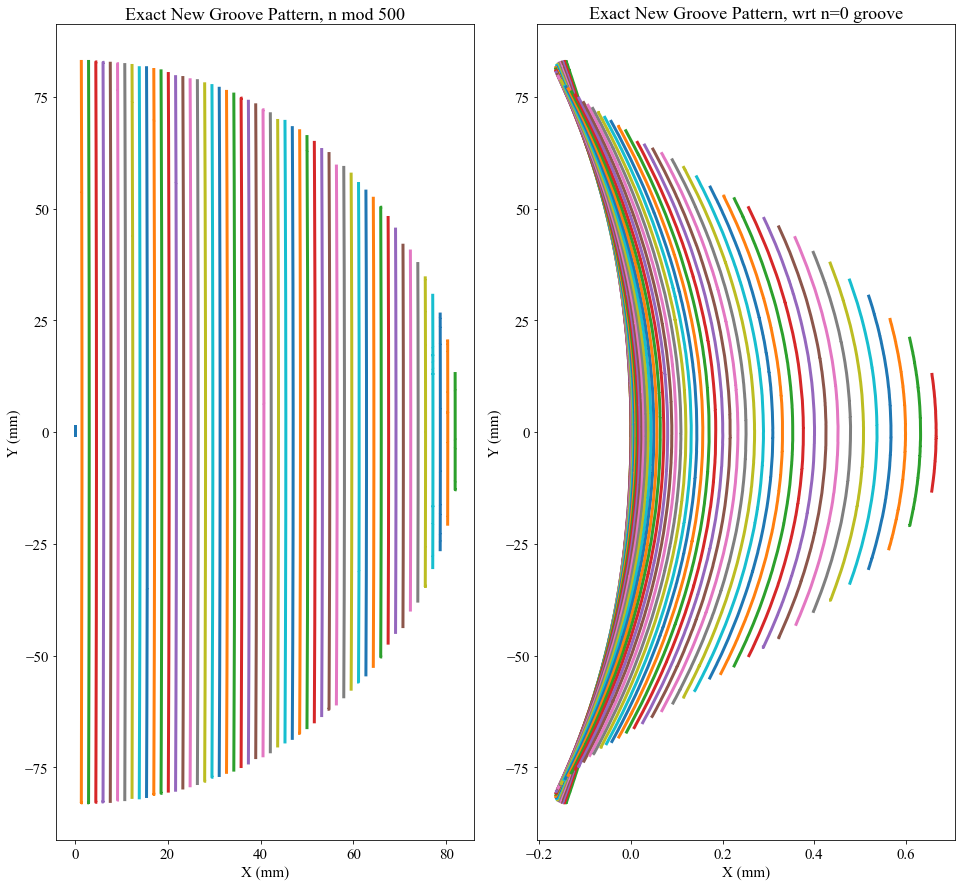}
   \end{tabular}
   \end{center}
   \caption[example] 
   { \label{new} 
(Left) – Half of the groove pattern on new grating, shown n mod 500. (Right) – Algebraically derived groove pattern for new grating solution with respect to the effective width. }
   \end{figure}

\section{FUTURE WORK AND CONCLUSION}
\label{sec:concl}
Though the blazed grating with zero order described in Section 2 has shown to be superior in minimizing the effects of astigmatism, it could likely be improved upon. Using the formalism of Noda et al.\cite{Noda74a,Noda74b}  and Wilkinson et al.\cite{Wilkinson} as inspiration, we will explore optimization strategies for aberration correction to maximize the impact of this new grating design.

Once a design is solidified, these grooves will be etched into a crystal silicon wafer via e-beam lithography. The next step is to transform these patterns into e-beam tool language for fabrication. The positions and widths will be written to a text file as a chain of paths in a format that is readable by the e-beam preparation software.\cite{Kruczek,Grise21}\footnote{See Kruczek et al., 2021 and Gris\'e et al., 2021 in this SPIE proceedings volume.} The process for fabricating the gratings so far has included choosing the materials and steps needed for fabrication and communicating with vendors to complete them. The primary difference between the laminar and blazed gratings during fabrication is their crystal orientation. The blazed grating has a challenging 1° blaze angle, something that would not be possible for conventional holographic ruling processes. Half of the light from the OAxFORTIS aperture is being blocked due to the off-axis design, so this blaze angle provides us a necessary factor of two in light gathering. So far, we have been sourcing silicon wafer substrates so they may then have an ellipsoid formed into them. The formed and polished wafers will later be given a nitride cap and will be spin-coated with an EBL-resist that will prepare them for the groove pattern to be etched into their surfaces.

Going into the rest of the year, we will continue efforts to optimize aberration minimization prior to the start of grating fabrication. Thus far we have demonstrated a precise method for deriving the pattern of variably-spaced grooves to be used on an elliptical diffraction grating. This grating, along with the other changes and technologies included in the redesign of FORTIS, will enable our science goals of demonstrating the scientific ability of multiobject spectroscopy in the far-UV. OAxFORTIS is designed to accomplish many including classifying far-UV spectra for the blue straggler population in the Globular Cluster M10, investigating low metallicity star formation in the Magellanic Bridge, and observing ionization stratification in the Cygnus Loop supernova remnant to unravel shock structures.

\newpage
\begin{center} {\bf ACKNOWLEDGEMENTS} \end{center}

This work is supported by NASA APRA grant NNX17AC26G to JHU  entitled, Rocket and Laboratory Experiments in Astrophysics -- Validation and Verification of the Next Generation FORTIS, and by a NASA SAT sub-award 80NSSC19K0450 to JHU entitled, Electron Beam Lithography Ruled Gratings for Future Ultraviolet/Optical Missions: High-Efficiency and Low-Scatter in the Vacuum Ultraviolet.

\bibliography{MC_SPIE21} 

\begin{thebibliography}{10}

\bibitem{McEntaffer:2013}
{McEntaffer}, R., {DeRoo}, C., {Schultz}, T., {Gantner}, B., {Tutt}, J.,
  {Holland}, A., {O'Dell}, S., {Gaskin}, J., {Kolodziejczak}, J., {Zhang},
  W.~W., {Chan}, K.-W., {Biskach}, M., {McClelland}, R., {Iazikov}, D., {Wang},
  X., and {Koecher}, L., ``{First results from a next-generation off-plane
  X-ray diffraction grating},'' {\em Experimental Astronomy}~{\bf 36},
  389--405 (Aug. 2013).

\bibitem{Grise:2020}
{Gris{\'e}}, F., {McEntaffer}, R., {Miles}, D., {McCurdy}, R., {Kruczek}, N.,
  {France}, K., {Fleming}, B., {Muslimov}, E., {Bouret}, J., {Caillat}, A., and
  {Chad}, E., ``{Opening the road to custom astronomical UV gratings},'' in
  [{\em American Astronomical Society Meeting Abstracts
  \#235}{\nolinebreak\hspace{0.1em}]},  {\em American Astronomical Society
  Meeting Abstracts} {\bf 235},  373.16 (Jan. 2020).

\bibitem{Beasley:2019}
{Beasley}, M., {McEntaffer}, R., and {Cunningham}, N., ``{Advances in
  aberration-correcting gratings using electron beam fabrication techniques},''
  in [{\em UV, X-Ray, and Gamma-Ray Space Instrumentation for Astronomy
  XXI}{\nolinebreak\hspace{0.1em}]},  {\em Society of Photo-Optical
  Instrumentation Engineers (SPIE) Conference Series} {\bf 11118},  1111816
  (Sept. 2019).

\bibitem{McCoy:2020}
{McCoy}, J.~A., {McEntaffer}, R.~L., and {Miles}, D.~M., ``{Extreme Ultraviolet
  and Soft X-Ray Diffraction Efficiency of a Blazed Reflection Grating
  Fabricated by Thermally Activated Selective Topography Equilibration},'' {\em
  \apj}~{\bf 891},  114 (Mar. 2020).

\bibitem{Fleming:2017}
{Fleming}, B., {Quijada}, M., {Hennessy}, J., {Egan}, A., {Del Hoyo}, J.,
  {Hicks}, B.~A., {Wiley}, J., {Kruczek}, N., {Erickson}, N., and {France}, K.,
  ``{Advanced environmentally resistant lithium fluoride mirror coatings for
  the next generation of broadband space observatories},'' {\em \ao}~{\bf 56},
  9941 (Dec. 2017).

\bibitem{Quijada:2017}
{Quijada}, M.~A., {del Hoyo}, J., {Boris}, D.~R., and {Walton}, S.~G.,
  ``{Improved mirror coatings for use in the Lyman Ultraviolet to enhance
  astronomical instrument capabilities},'' in [{\em Society of Photo-Optical
  Instrumentation Engineers (SPIE) Conference
  Series}{\nolinebreak\hspace{0.1em}]},  {\em Society of Photo-Optical
  Instrumentation Engineers (SPIE) Conference Series} {\bf 10398},  103980Z
  (Sept. 2017).

\bibitem{Kim:2021}
Kim, K., Kutyrev, A.~S., Li, M.~J., and Greenhouse, M.~A., ``Actuation force
  analysis and design optimization of microshutter array by numerical
  simulation method,'' {\em Engineering Research Express}~{\bf 3},  015007 (jan
  2021).

\bibitem{Kutyrev:2020}
{Kutyrev}, A.~S., {Greenhouse}, M., {Li}, M.~J., {Kim}, K., {Brekosky}, R.,
  {McCandliss}, S., {Costen}, N., and {Wang}, F., ``{Programmable microshutter
  selection masks in application to UV spectroscopy},'' in [{\em Society of
  Photo-Optical Instrumentation Engineers (SPIE) Conference
  Series}{\nolinebreak\hspace{0.1em}]},  {\em Society of Photo-Optical
  Instrumentation Engineers (SPIE) Conference Series} {\bf 11443},  114431D
  (Dec. 2020).

\bibitem{McCandliss10}
McCandliss, S., Fleming, B., Kaiser, M., Kruk, J., Feldman, P., Kutyrev, A.,
  Li, M., Goodwin, P., Rapchun, D., Lyness, E., Brown, A., Moseley, H.,
  Siegmund, O., and Vallerga, J., ``Fabrication of {FORTIS},'' {\em SPIE}~{\bf
  7732},  773202 1--12 (2010).

\bibitem{Noda74a}
Noda, H., Namioka, T., and Seya, M., ``Geometric theory of the grating,'' {\em
  Journal of the Optical Society of America}~{\bf 64},  1031--1036 (1974a).

\bibitem{Noda74b}
Noda, H., Namioka, T., and Seya, M., ``Ray tracing through holographic
  gratings,'' {\em Journal of the Optical Society of America}~{\bf 64},
  1037--1045 (1974b).

\bibitem{Wilkinson}
Wilkinson, E., Indebetouw, R., and Beasley, M., ``Technique for narrow-band
  imaging in the far ultraviolet based on aberration-corrected holographic
  gratings,'' {\em Journal of the Optical Society of America}~{\bf 40},
  3244--3255 (2001).

\bibitem{Tsonev}
Tsonev, L. and Popov, E., ``Focal images formed by a concave holographic
  grating. a corroborative investigation using three different techniques,''
  {\em Journal of Modern Optics}~{\bf 39},  1749--1760 (1992).

\bibitem{Kruczek}
Kruczek, N. et~al., ``Performance of anisotropically-etched gratings in the
  extreme and far ultraviolet bandpasses,'' (in prep).

\bibitem{Grise21}
Gris{\'e}, F. et~al., ``Fabrication of custom astronomical gratings for the
  extreme and far ultraviolet bandpasses,'' (in prep).

\end{thebibliography}
\bibliographystyle{spiebib} 

\end{document}